\documentclass[runningheads]{llncs}
\usepackage{graphicx}
\usepackage[usenames, dvipsnames]{color}
\begin{document}
\title{Reversibility in space, time, and computation: the case of underwater acoustic communications\thanks{This publication has emanated from research supported in part by a research grant from Science Foundation Ireland (SFI) and is co-funded under the European Regional Development Fund under Grant Number 13/RC/2077. The project has received funding from the European Union’s Horizon 2020 research and innovation programme under the Marie Skłodowska-Curie grant agreement No 713567 and was partially supported by the COST Action IC1405.}}
\subtitle{Work in Progress Report}
\titlerunning{Reversibility and Underwater Acoustic Communications}
%
\author{Harun Siljak\orcidID{0000-0003-1371-2683}}
\authorrunning{H. Siljak}
%
\institute{CONNECT Centre, Trinity College Dublin, Ireland\\
\email{siljakh@tcd.ie}}
\maketitle              
\begin{abstract}
Time reversal of waves has been successfully used in communications, sensing and imaging for decades. The application in underwater acoustic communications is of our special interest, as it puts together a reversible process (allowing a reversible software or hardware realisation) and a reversible medium (allowing a reversible model of the environment). This work in progress report addresses the issues of modelling, analysis and implementation of acoustic time reversal from the reversible computation perspective. We show the potential of using reversible cellular automata for modelling and quantification of reversibility in the time reversal communication process. Then we present an implementation of time reversal hardware based on reversible circuits.
\end{abstract}
\section{Introduction}

The idea of wave time reversal has been considered for decades: among other references, we can find an early mention in Rolf Landauer's work \cite{landauer1960parametric}. The theory and practice of modern time reversal of waves stems from Mathias Fink's idea of time reversal mirrors \cite{fink1992time}. While the first theoretical and practical results came from the case of sound waves (acoustics), the concept was translated in electromagnetic domain as well, through applications in optics \cite{popoff2011exploiting} and radio technology \cite{lerosey2004time}. The particular scenario we are considering here is the case of underwater acoustic communications (UAC), an application where the poor electromagnetic wave propagation makes sound waves the best solution.

Fluid dynamics (motion of liquids and gases) using reversible cellular automata (RCA)  has been discussed extensively in the past \cite{frisch1986lattice,margolus1986cellular}. However, this century has seen only a few applications of RCA to macro-scale engineering problems such as acoustic underwater sensing \cite{mckerrow2001simulating}. Similarly, proofs of concept for hardware and software implementations of reversible digital signal processing exist \cite{de2010reversible}, and we yet have to see it applied. This work will combine these underutilised methods of modelling and hardware implementation and relate them to time reversal, another physical form of reversibility. The work is fully practical in nature, as it is conducted on the UAC use case. 

The style of exposition is tailored to give the first introduction to time reversal of waves to reversible computation community. Linking the two fields in both analytic (RCA) and synthetic sense (reversible hardware) is the main contribution of this paper: in future publications we will present the results. The wide approach taken, ranging from physical phenomena to cellular automata and reversible hardware is intended to show the role of different aspects of reversibility in a single practical application. We first introduce time reversal and its application in UAC, followed by a section motivating the use of RCA in modelling and quantification of the time reversal process. Second, we present a reversible hardware solution for the time reversal processing chain and a brief set of conclusions and future work pointers.

\section{State of the Art: Time Reversal}

\begin{figure}[t]
\centering{\includegraphics[width=\textwidth]{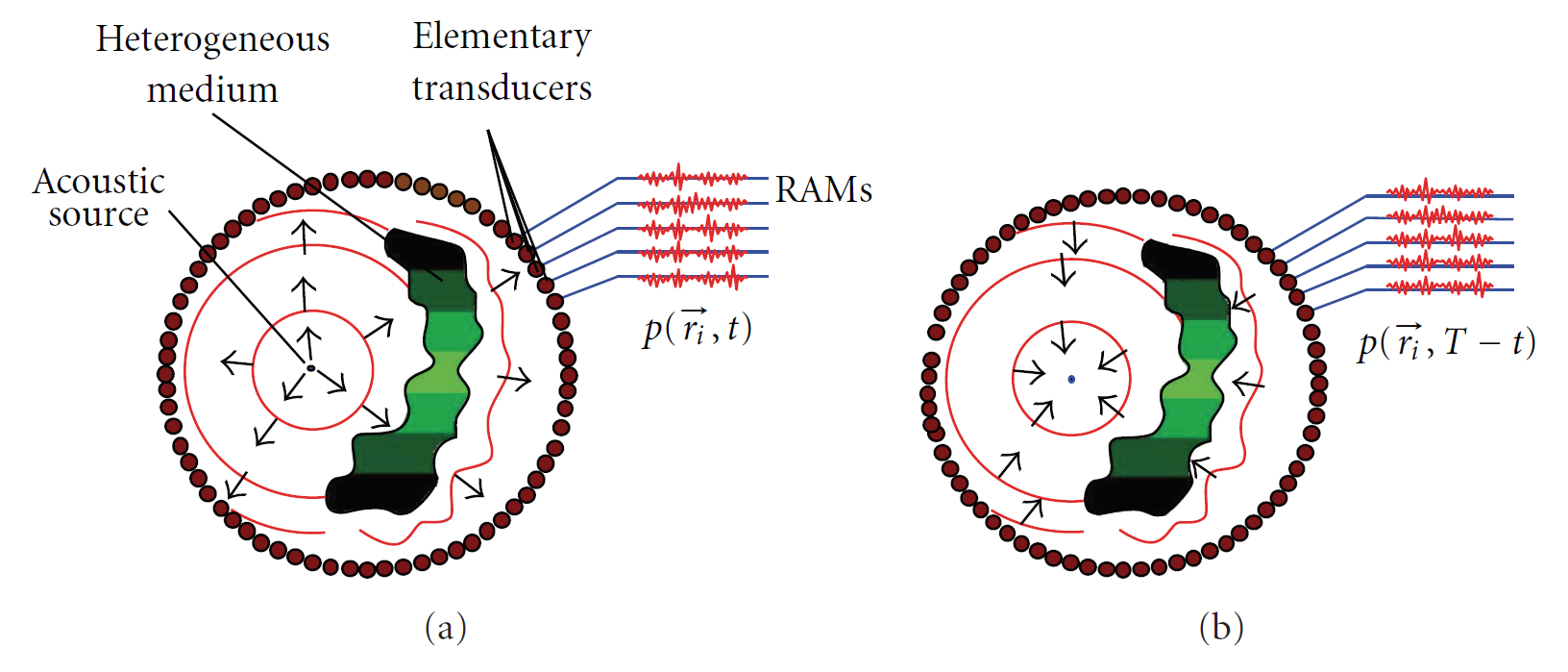}
\caption{The time reversal mechanism in a cavity (from \cite{lemoult2011time})} \label{fink1}}
\end{figure}

The concept of (acoustic) time reversal is illustrated in Fig.~\ref{fink1} \cite{fink1992time}. If we place a sound source in a heterogeneous medium within a cavity and let it emit a pulse, this pulse will travel through the medium and reach an array of transducers\footnote{From the electronic point of view, these elements are piezo transducers capable of converting mechanical to electrical energy while operating as receivers and the opposite while operating as transmitters. From the communications standpoint, they are transceivers, and from the everyday standpoint they are microphones/speakers.} placed on the cavity walls (Fig.~\ref{fink1}(a)). If we emit what the transducers have received in a reversed time order (Fig. ~\ref{fink1}(b)), we will get a sound wave resembling an echo. However, unlike an echo, this sound wave will not disperse in the cavity, but be focused instead, converging at the point where the original wave was generated, at the acoustic source.

\begin{figure}[t]
\includegraphics[width=\textwidth]{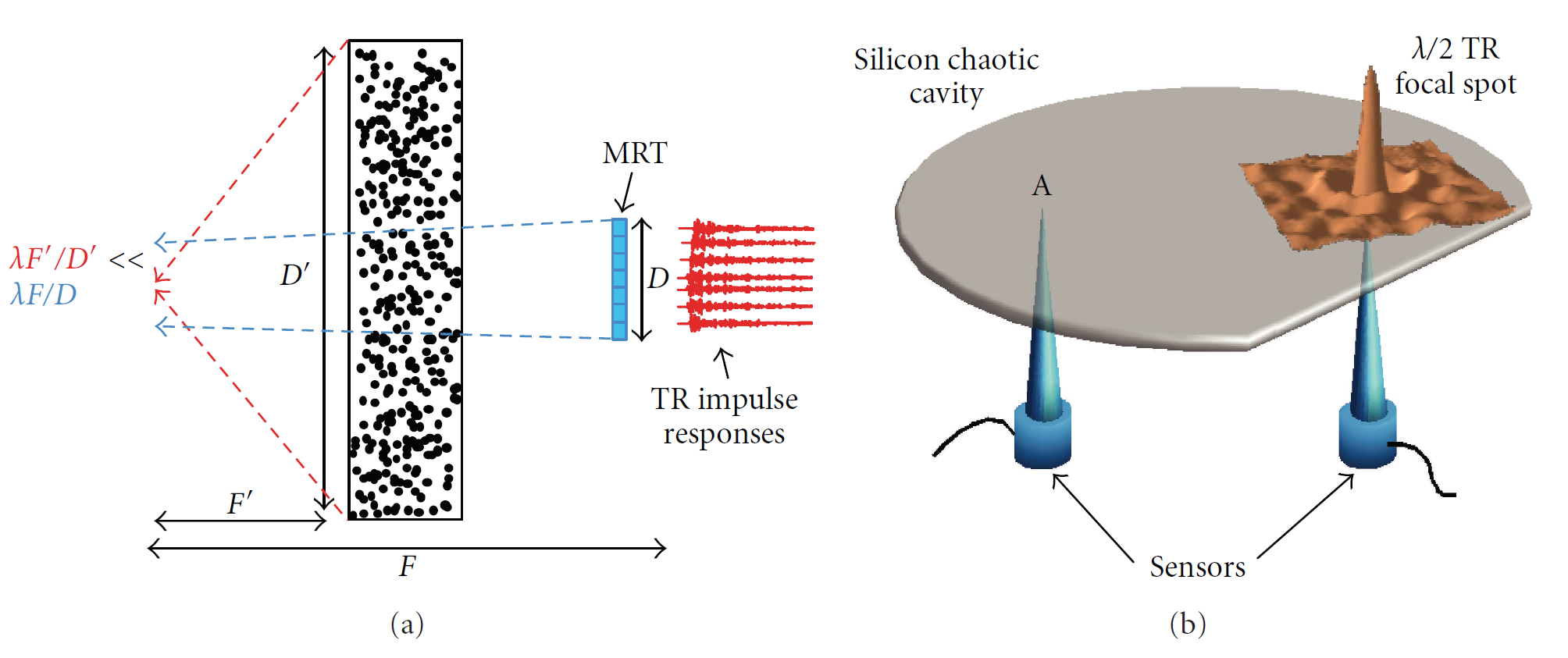}
\caption{(a) Localised time reversal mirror with a complex propagation medium (b) A simplified reversal scheme with a 3-D focal spot visualisation (from \cite{lemoult2011time})} \label{fink2}
\end{figure}

Covering the whole cavity with transceivers is not feasible: not only does it ask for a large number of transceivers, but sometimes the system is deployed in (partially) open space, not a cavity. Hence, the option of a localised time reversal mirror (TRM) has to be considered: only a few transceivers co-located in a single position that use the multipath effects resulting from multiple reflections of the emitted sound wave on the scatterers in the environment. As it turns out, it is possible to have the time reversal effect and a coherent pulse at the original source if the environment is complex enough (Fig.~\ref{fink2}(a) illustrates such an experiment). This counter-intuitive result relies on the effect of ergodicity inherent to ray-chaotic systems \cite{draeger1999one}: the wave will pass through every point in space eventually and collect all the environment information on the way to the mirror.

The application of this concept to communications is straightforward: the wave, when returned to the original sender may convey information from the receiver (TRM) and it will be focused only at the location of the original source (preventing both eavesdropping and interference at other locations). While there are other applications as well (e.g. localisation, imaging) we focus on the communications aspect as it allows us to introduce multiple transmitters and receivers (multiple inputs, multiple outputs, MIMO) and analyse the effects of (irreversible) signal interference in the (reversible) model.

\section{Modelling and Quantification}

Time reversal in the UAC setting is an example of a reversible process in a nominally reversible environment. While dynamics of water (or any fluid for that purpose) subject to sound waves, streams, waves and other motions are inherently reversible, most of the sources of the water dynamics cannot be reversed: e.g. we cannot reverse the Gulf stream or a school of fish even though their motion and the effect on water is in fact reversible. Hence, even though it would rarely be completely reversed, the model for UAC should be reversible.

\begin{figure}[t]
\centering{\includegraphics[width=0.8\textwidth]{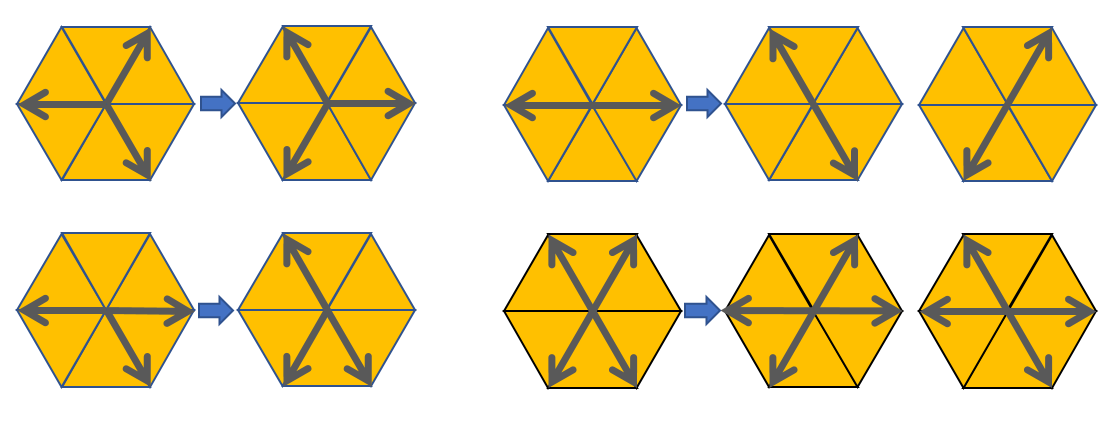}
\caption{Collision rules for FHP gas} \label{fhp}}
\end{figure}

RCA give us such an option through the lattice gas models \cite{succi2001lattice}: cellular automata obeying the laws of fluid dynamics described by the Navier-Stokes equation. One such model, the celebrated FHP (Frisch-Hasslacher-Pomeau) lattice gas \cite{frisch1986lattice} has had several improvements after its original statement in 1986 \cite{wolf2004lattice}, but its basic form is simple and yet following the Navier-Stokes equations exactly.  This is a model defined on a hexagonal grid through a set of rules of particle collision shown in Fig.~\ref{fhp}. The model can be interpreted as an RCA via partitioning approaches\footnote{Partitioning of cellular automata is an approach rules are applied to blocks of cells and the blocks change in successive time steps. Different approaches exist, depending on the grid shape, e.g. Margolus neighbourhood for square grids, and Star of David and Q*Bert neighbourhoods for hexagonal grids.} \cite{toffoli1987cellular}, but the randomness of transitions when collisions include more possible outcomes (as seen in the figure) has to be taken into account.

The FHP lattice gas provides us a two-dimensional model for UAC, easily implementable in software and capturing the necessary properties of the reversible medium. It is not a novel idea to use a lattice gas to model water, but neither acoustic underwater communications or time reversal of waves have been observed through this lens before. As already noted, however, time reversal is not going to be conducted by running the cellular automaton backwards in time, as that would reverse parts of the environmental flow we usually have no influence over. The acoustic time reversal is performed the same way as in real systems, by time reversing the signal received at the time reversal mirror.

The model we observe consists of the original source (transmitter) which causes the spread of an acoustic wave, the original sink (receiver) waiting for the wave to reach it, as well as scatterers and constant flows (streams) in the environment. The constant stream and the loss of information caused by some wave components never reaching the sink will result in an imperfect reversal at the original source when the roles are switched (i.e. when the time reversal mirror returns the wave). If we measure the power returned, we will have a directivity pattern (focal point) similar to the one in Fig.~\ref{fink2}(b). 

The amplitude of the peak will fluctuate based on the location of the original source and may serve as a metric: a measure of reversibility. If we move the source over the whole surface of the model and measure this metric (whose analogue in quantum reversibility studies is fidelity or Loschmidt Echo \cite{taddese2009chaotic}) we obtain a heatmap of the surface with respect to the quality of time reversal. In the context of time reversal studies, it is used as a measure of the quality of communication, but in a more general context it can measure reversibility of a cellular automaton. The functionality of the model increases if we observe several transceivers distributed over the area (e.g. underwater vehicles communicating with a central communication node) and/or allow motion of transceivers. The complexity of the model increases as well, and the reversibility metrics become a measure of interference. This is the first part of our ongoing work, as we investigate the effects contributing to communication quality loss in the FHP model for UAC.

\section{Reversible Hardware Implementation}

\begin{figure}[t]
\centering{\includegraphics[width=\textwidth]{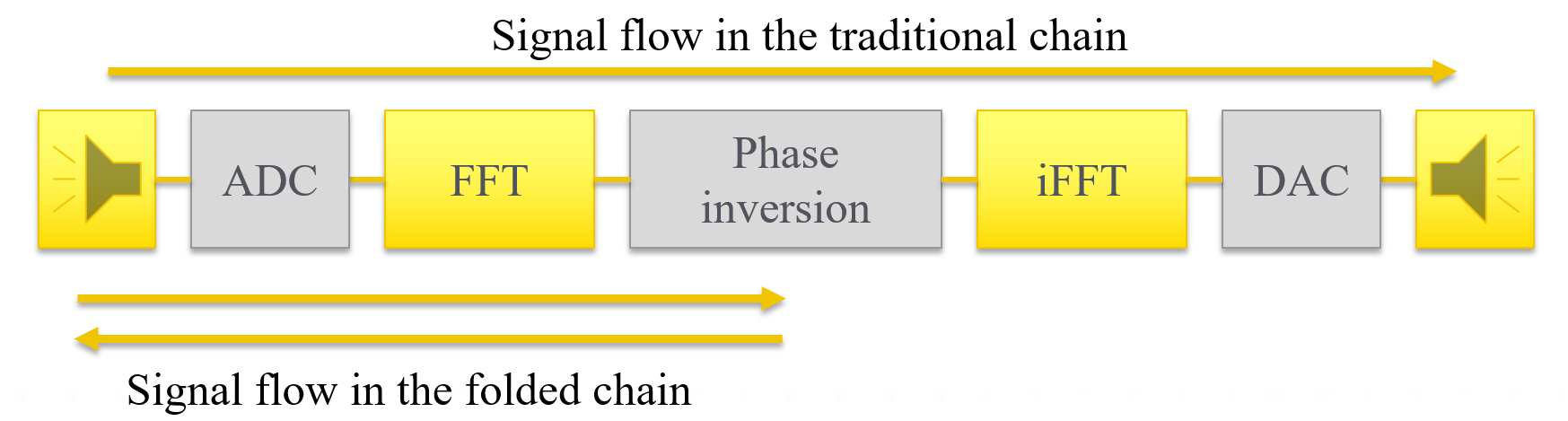}
\caption{The reversible hardware scheme for acoustic time reversal} \label{fig1}}
\end{figure}

The reversibility in time of the communication scheme we use and the reversibility in space of the medium both suggest that the reversibility in computation should exist as well.  Fig.~\ref{fig1} gives an overview of a reversible architecture we are proposing, which consists of speakers/microphones, AD/DA (analogue to digital/digital to analogue) converters, Fast Fourier Transform (FFT) blocks and a phase conjugation block. This architecture is already the one used in wave time reversal--here we interpret it in terms of reversible hardware. 

From the electromechanical point of view, a microphone and a speaker are the same device, running on the same physical principle, which makes the two ends of the scheme equivalent. The next element, the AD converter on one and DA converter on the other end are traditionally made in an irreversible fashion as the signal in traditional circuits flows unidirectionally. However, one of the first categories in the international patent classification of AD/DA converters is H03M1/02: Reversible analogue/digital converters. There has been a significant number of designs proposed to allow bidirectional AD/DA conversion. With that in mind, we may consider this step to be reversible as well. 

The signal received is manipulated in the Fourier (frequency) domain by conjugation (change of the sign of the complex image's phase) as conjugation in frequency domain results in time reversal in time domain. This asks for a chain of transform, manipulation and inverse transform so the new time domain signal can be emitted. All elements in this chain are inherently reversible. Reversibility of the Fourier transform has been long utilised, and reversible software and circuit implementations of its commonly used computational scheme, FFT have been proposed \cite{li2004reversible,skoneczny2008reversible,yokoyama2008principles}. Hence, the FFT block can be considered reversible, and the Inverse Fast Fourier Transfom (IFFT) is just the FFT block with the reversed flow. Finally, the phase reversal is simply changing the sign of the half of the outputs coming from the FFT block, as the whole set of outputs comprises of phase and amplitude of the signal in frequency domain. Changing the sign is the straightforwardly reversible action of subtraction from zero or simple complementation of the number, and as such has been solved already in the study of reversible arithmetical logic unit \cite{thomsen2010reversible}.

Once we have determined the reversibility of the scheme, we note its symmetry as well. If we fold the structure in the middle (at the conjugation block), the same hardware can be used both to propagate the inputs and the outputs. While the particular details of circuit implementation are left for future work, where details of the additional circuitry will be addressed as well, we have here presented this scheme as a proof of concept, a reversible signal processing scheme convenient for an implementation in reversible hardware. That is the second part of our ongoing work. 



\section{Conclusion}

In this work in progress report, we have presented the potential of reversible computation for time reversal in UAC. Future work will focus on both the modelling prospects using RCA and the reversible circuit implementation of the time reversal hardware. While going into more detail to cover all the practical issues of it, future work also needs to address the appropriateness of the same or similar approach to the question of time reversal in optics and radio wave domain. The integration of reversible computation with physical time reversal in this context opens a general discussion on the relationship of different interpretations of reversibility and new venues for reversible computation.

%
%
%

%
%
%
%
%
\end{document}